\documentclass[12pt]{iopart}
\usepackage{graphicx}
\usepackage{pdflscape}
\begin{document}

\title{A General Bidirectional Controlled/Uncontrolled Quantum Teleportation Protocol}

\author{Moein Sarvaghad-Moghaddam1$^{1}$ \&
         Negin Fatahi$^{2*}$            
        }

\address{$^{1}$Young Researchers and Elite Club, Mashhad Branch, Islamic Azad University, Mashhad, Iran. .}
\address{$^{2}$Department of Physics, Kermanshah Branch, Islamic Azad University, Kermanshah,Iran.}
\ead{$^{*}$n83fatahi@yahoo.com} \vspace{10pt}

\begin{indented}
\item[]
\end{indented}

\begin{abstract}
In this paper, different from the other existing methods that only the special and limited numbers of qubits transmit between Alice and Bob, a general method is proposed to implement Bidirectional Quantum Controlled/uncontrolled Teleportation (BQCT/BQT) that Alice and Bob can transmit each type of arbitrary n and m-qubits to each other, simultaneously. This protocol is based on Controlled-NOT (CNOT) operation, appropriate single-qubit unitary operations and single-qubit measurement in the Z-basis and X-basis. Also, quantum channel can be created by using Hadamard and CNOT operations. One of the advantages of this method is using the single-qubit measurements which are more efficient than two-qubit measurements.
\end{abstract}

\vspace{2pc} \noindent{\it Keywords}: Bidirectional quantum teleportation, n-qubit state, Bidirectional quantum controlled teleportation, entanglement state.

\section{Introduction}
Quantum teleportation (QT) is one of the communication protocols in Quantum information theory where an unknown quantum state can be transmitted to a receiver by using the shared entanglement state as quantum channel and some auxiliary classical communications \cite{Nielsen}. \\
Rapidly, safe and secret communication has been caused QT to become a very interesting topic of researches and plays an important role in many applications in recent years [2-8]. In 1993, Bennett et.al.\cite{Bennett}, introduced idea of QT by using an Einstein-Podolsky-Rosen (EPR) pair as a quantum channel. After that, modified versions of QT are introduced by using EPR pair, Greenberger- Horne-Zeilinger (GHZ) state, W state and other entangled states as a quantum channel [9-14]. \\
   In 1993, Karlsson et al.\cite{Karlsson}, proposed another version of QT named as Controlled Quantum Teleportation (CQT). This protocol is same as QT with presence of third user (Charlie) as controller or supervisor. Later, in [16-23] several protocols of CQT were proposed with one or more controller. 
   In 2013, Zha et al. \cite{Zha} demonstrated a new version of QT named as Bidirectional Controlled Quantum Teleportation (BCQT) via five-qubit cluster state as quantum channel. In BCQT or BQT protocol, both authorized users transmit unknown quantum states to each other, simultaneously. In \cite{Yan2,Sun}, BCQT protocols are proposed with various channels such as six-qubit cluster state, six-qubit entangled state and five-qubit composite. Also, Chen \cite{Chen}, Hassanpour \cite{Hassanpour}, Hong \cite{Hong} and Sang \cite{Sang}, proposed various BCQT protocols by using channels of six-qubit genuine, six-qubit entangled state and seven-qubit entangled state, respectively. In proposed protocol in \cite{Hong,Sang} Bob and Alice can transmit arbitrary two and one qubit together by using a seven-qubit entangled state as quantum channel. In \cite{Hassanpour2}, a pure EPR state can be transmitted by Alice and Bob to each other by using six-qubit GHZ as quantum channel. In proposed protocol by Li et al. \cite{Li3}, an arbitrary two and one- qubit  can be transmitted by Alice and Bob via six-qubit cluster state simultaneously. In 2017,  Sadeghi Zadeh et al. \cite{Zadeh} proposed a BQT protocol which the users can transmit an arbitrary two-qubit state to each other by using eight-qubit quantum channel.
   In this paper, a new general protocol of Bidirectional Quantum Teleportation (BQT) and Bidirectional Quantum Controlled Teleportation (BQCT) are proposed for transmitting each number of unknown qubits between Alice and Bob with supervisor Charlie or without it. This method can covers the transmitting of each arbitrary type of qubits such as entanglement states or pure states. In this protocol, users only perform single-qubit measurements.\\
   The remainder of this paper is organized as follows. In Section 2, proposed method for general BQT protocol is introduced  in details. In Section 3, the comparison between reported channels in previous works and our method is discussed. Finally, Section 4 concludes the paper.
   
\section{General BQT protocol }   
In this section, firstly the general  Bidirectional Quantum Teleportation (BQT) protocol is proposed for transmitting n and m-qubits between Alice and Bob. Then this proposed method  generalized for transmitting the n- qubit entanglement state between one or two users to each other. Finally, a general method is proposed for Bidirectional Controlled Quantum Teleportation (BQCT) protocol with presence of Charlie as a controller or supervisor.
\subsection{Transmitting the n and m qubits state}
In this section, a general method for BQT protocol is proposed. By using this protocol Alice and Bob can transmit an arbitrary n and m-qubit state to each other. The general schematic of this protocol is shown in Fig. 1.

\begin{figure}[h!]
\center
  \includegraphics[width=1\textwidth]{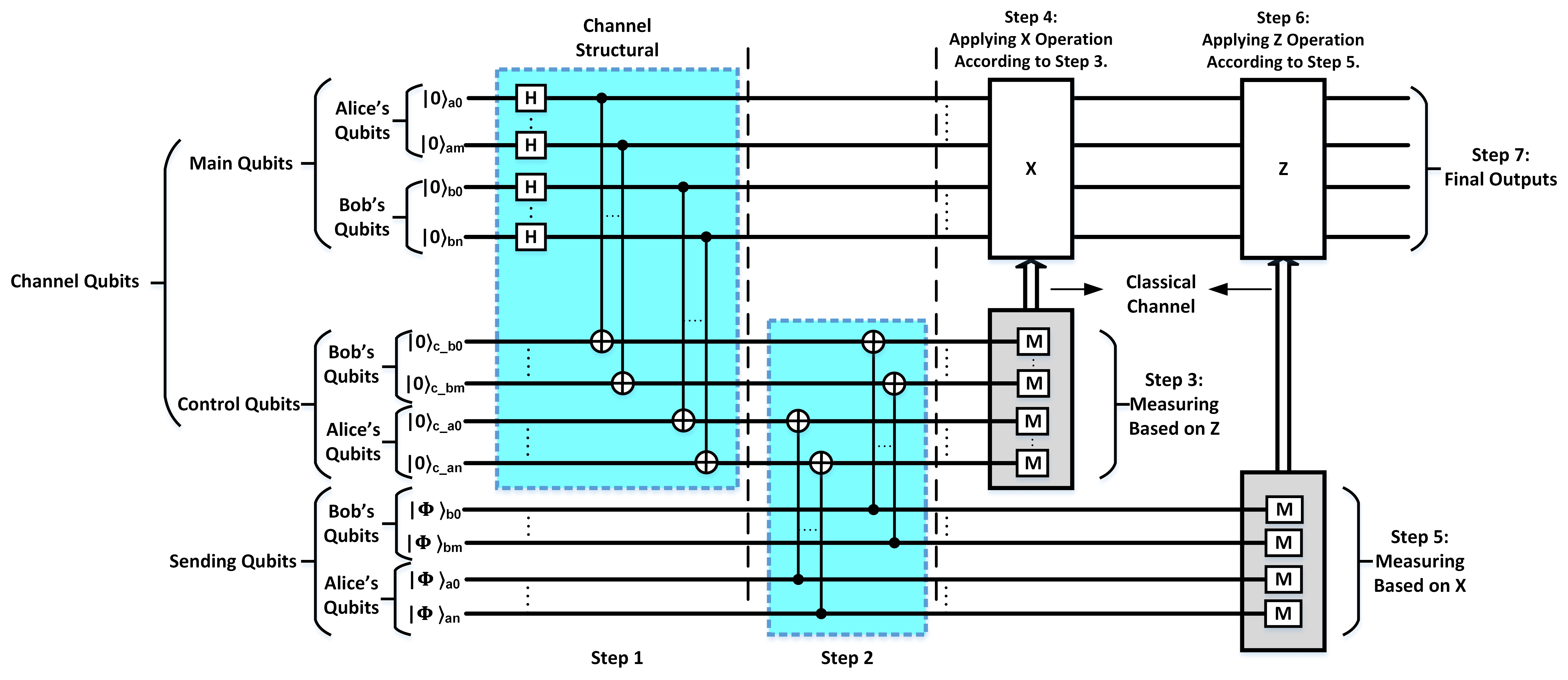}
 \caption{General schematic of proposed BQT protocol for transmitting the arbitrary n and m-qubit states between Alice and Bob.}
 \label{1}
\end{figure}

Consider the n-qubit general state that Alice wants to teleport to Bob and the m-qubit general state that Bob wants to teleport to Alice. This states are given by 

\begin{equation}
\vert\phi_a>=\sum_{i=0}^{2^{n}-1}\alpha_{i}\vert i> \;\;\;\;\alpha_{i}\in c_z, \;\;\;\;\sum_{i=0}^{2^{n}-1}\alpha_{i}^2=1
\end{equation}

\begin{equation}
\vert\phi_b >=\sum_{i=0}^{2^{m}-1}b_{i}\vert i> \;\;\;\;b_{i}\in c_z, \;\;\;\;\sum_{i=0}^{2^{n}-1}b_{i}^2=1 
\end{equation}

Where variable i is binary number. As shown in Fig. 1, the qubits that used in channel are divided into two parts: Main qubits and Control qubits. Duty of main qubits is creating the number of required qubits for sending information between Alice and Bob ($|0...0>_{a} |0...0>_b $ that $ a,b$ stand to belong to Alice and Bob respectively). Sending qubits show the number of qubits transmitted by Alice and Bob to each other. The numbers of Main qubits are dependent on the sum of the Sending qubits by Alice and Bob. If the number of Sending qubits by Alice is $n$ and the number of Sending qubits by Bob is $m$, so the number of Main qubits is $m+n$. The number of Control qubits are same as Main qubits and n-qubit is belong to Alice and m qubit belong to Bob. In the other words, the number of Control qubits for each of Alice and Bob are equivalent to the number of the Sending qubits. These additional qubits are used for communicating with Sending qubits. The placing of Control qubits is arbitrary. We place them after than Main qubits. The protocol consists the following steps:
\\
\textbf{Step1:}\\
In this step, quantum channel is created. As shown in Fig. 1, the first Main qubits and Control qubits are created as stated in above. Also, the arrangement of the Main qubits are inverses of Sending qubits. But, the arrangement of the Control qubits is same as Sending qubits. For example, if the arrangement of Sending qubits are as $\vert\phi_{bm}> \vert\phi_{an}>$ then, the arrangement of Main qubits and Control qubits are $\vert0...0>_{am}\vert0...0>_{bn}\vert0...0>_{c_{bm}}\vert0...0>_{c_{an}}$ where $a,b$ and $c$ stand to belong to Alice, Bob and Control qubits, respectively. Also, the index n and m is the number of qubits that transferred by Alice and Bob. Then, Hadamard gate is applied to the Main qubits and create a superposition from this qubits and after that CNOT gate is applied, respectively. So that, Main qubits are used as Control Lines and Control qubits are used as target lines. For example, the first qubit $|0>_{a0}$  (that belongs to Alice) is used as Control line for target line $|0>_{c_{b0}}$ (the first Control qubit that belongs to Bob) and so on. So, generally same states of Main qubits are multiplied as a tensor product with control qubits. Thus, the general state can be written as:
\begin{equation}
\frac{1}{\sqrt{n}}\left( \sum_{i=0}^{2^{n+m}-1}\vert i>_{ab}\vert i>_{c_{b}}\vert i>_{c_{a}}\right)\otimes\vert \phi>_b \vert \phi>_a
\end{equation} 
where, $n$ and $m$ are the number of Sending qubits by Alice and Bob, respectively. Also, the index $a$ and $b$ belong to Alice and Bob and $c_b ,c_a$ show the Control qubits belong to Bob and Alice. $|\phi>_b$ and $|\phi>_a$ denote Sending qubits by Bob and Alice.
\\
\textbf{Step2:}\\
Controll NOT operator is applied by Alice and Bob so that, Sending qubits are as Control lines and Control qubits are as target lines. For example, consider the Sending qubits belong to Bob and Alice as $|\phi>_b={|\phi>_{b0},|\phi>_{b1},...,|\phi>_{bn}}$, and $|\phi>_a={|\phi>_{a0},|\phi>_{a1},...,|\phi>_{am}}$, for instance, CNOT gate can be used by Bob (Alice) as $|\phi>_{b0}(|\phi>_{a0})$ as control line and line of $c_{b0} (c_{a0})$ is used as target line and so on.\\
\textbf{Step 3:}\\
In this step, Alice and Bob measure the single qubit of the control qubits in the Z-basis. The unmeasured qubits collapse into the one of the $2^{(n+m)}$ possible states with equal probability.\\
\textbf{Step 4:}\\
 Alice and Bob  share the measurement results that obtained from previous step, on classical channel. Then, they apply $X$ unitary operator to their unmeasured qubits. If for each qubit in control line, $1$ is obtained then, the $X$ unitary operation applied to equivalent qubit in the main qubits. For example, if $c_{a0}$  be $1$, Then $X$ operator is applied to equivalent qubit in Main qubits i.e. $a0$ line.\\
\textbf{ Step 5:}\\
In this step, Alice and Bob  measure the single-qubit of Sending qubits in the $X$-basis. The unmeasured qubits collapse into one of the $2^{(n+m)}$ possible states with equal probability.\\
\textbf{Step 6:}\\
After that the users tell their measurement results to each other, they apply Z unitary operator to their unmeasured qubits. After that, if for each Sending qubits, the value of one is obtained then, for the equivalent qubit in the Main qubits, Z unitary operator is applied. For example, if  after that  measurement the value of $|\phi>_{a0}$  equal $|->$, then, $Z$ operator is applied to equivalent qubit in the Main qubits i.e. $a0$ line. Finally, all of the states will be in the same form.\\
\textbf{Step 7:}\\
Users reconstruct the $n$ and $m$-qubit states and the BQT is successfully finished.\\
See the following example in order to explain this method in more details.\\
\textbf{Example:}\\
Consider a two-qubit state that Alice wants to teleport to Bob and a one-qubit state that Bob want to teleport to Alice, that are given by: \\

\begin{equation}
\vert \phi>_{A_0 A_1}=\alpha_{00}|00> +\alpha_{01}|01>+\alpha_{10}|10>+\alpha_{11}|11>
\end{equation}
\begin{equation}
\vert \phi>_{B_0}=b_{0}|0> +b_{1}|1>
\end{equation}
This protocol consists of the following steps:\\
\textbf{Step 1:} For creating channel, two qubit belong to Bob and one qubit belong to Alice are considered as Main qubits. Then, two qubit for Alice and one qubit for Bob are considered as Control qubits. So, the number of qubits is as following:
\begin{equation}
\vert 000>_{b0b1a0}\vert 000>_{c_{a0}c_{a1}c_{b0}}
\end{equation}
By applying the Hadamard gate to the Main qubits we have:
\begin{equation}
\frac{1}{\sqrt{8}}(\vert 000>+\vert 001>+\vert 010>+\vert 100>+\vert 011>+\vert 110>+
\end{equation}
$$\vert 101>+\vert 111> )_{b0b1a0}\vert 000>_{c_{a0}c_{a1}c_{b0}}$$

Then, by applying CNOT gate to the Control qubits, the structure of the channel is created as the following:
\begin{equation}
|\psi>_{(b0)(b1)(a0)(c_{a0})(c_{a1})(c_{b0})}=\frac{1}{\sqrt{8}}(|000000>+|001001>+|010010> \end{equation}
$$
+|011011>+|100100>+|101101>+|110110>+|111111> 
$$
The state of the whole system can be expressed as :

\begin{equation}
|\phi>_{(b0)(b1)(a0)(c_{a0})(c_{a1})(c_{b0})A0A1B0}=|\psi>_{(b0)(b1)(a0)(c_{a0})(c_{a1})(c_{b0})} \otimes 
|\Phi>_{(A0 A1)}\otimes |\Phi>_{(B0)}
\end{equation}
\textbf{Step 2:} In this step,  Alice and Bob applied Controlled NOT (CNOT) operator so that $A0, A1$ and $B0$ as control line and $c_{a0},c_{a1}$ and $c_{b0}$ (Control Qubits ) as target line, respectively. Then the state of the whole system is shown as:
\begin{equation}
|\phi>_{(b0)(a0)(a1)(c_{a0})(c_{a1})(c_{b0})A0A1B0}=
\end{equation}
$$
\frac{1}{\sqrt{8}}[\alpha_{00} b_{0} (|000>|000>+|001>|001>+|010>|010>+|011>|011>+|100>|100>$$
$$+|101>|101>+|110>|110>+|111>|111>)|000>+$$
$$\alpha_{00} b_{1}(|000>|001>+|001>|000>+|010>|011>+|011>|010>+|100>|101> $$
$$|101>|100>+|110>|111>+|111>|110>)|001>+ $$
$$\alpha_{01} b_{0}(|000>|010>+|001>|011>+|010>|000>+|011>|001>+|100>|110>$$
$$+ |101>|111>+|110>|100>+|111>|101>)|010>+ $$
$$\alpha_{01} b_1 (|000>|011>+|001>|010>+|010>|001>+|011>|000>+|100>|111>$$
$$+ |101>|110>+|110>|101>+|111>|100>)|011>+ $$
$$\alpha_{10} b_0 (|000>|100>+|001>|101>+|010>|110>+|011>|111>+|100>|000>$$
$$+|101>|001>+|110>|010>+|111>|011>)|100>+$$
$$\alpha_{10} b_1 (|000>|101>+|001>|100>+|010>|111>+|011>|110>+|100>|001>$$
$$+|101>|000>+|110>|011>+|111>|010>)|101>+$$
$$ \alpha_{11} b_0 (|000>|110>+|001>|111>+|010>|100>+|011>|101>+|100>|010>$$
$$+|101>|011>+|110>|000>+|111>|001>)|110>+$$
$$\alpha_{11} b_1 (|000>|111>+|001>|001>+|010>|101>+|011>|100>+|100>|011>$$
$$+|101>|010>+|110>|001>+|111>|000>)|111>]$$

\textbf{Step3:} Single-qubit measurement is applied on qubits $c_{a0},c_{a1}$ and $c_{b0}$ (Control Qubits) by Alice and Bob, respectively. As shown in Table 1, the unmeasured qubits collapse into the one of the 8 possible states with equal probability.

\begin{table}[h!]
\caption{measurement results based on Z done by Alice and Bob on Control Qubits}
\label{tab:1}
\begin{tabular}{|c|c|c|}
\hline
Alice's & Bob's & The collapsed state of qubits \\
Result & Result & $(b0)(b1)(a0)A0A1B0$ encoding algorithm\\
\hline
00 & 0 &  $\alpha_{00} b_0|000>|000>+\alpha_{00} b_1 |001>|001>+\alpha_{01} b_0 |010>|010>+\alpha_{01} b_1|011>|011> $ \\
& & $+\alpha_{10} b_0 |100>|100>+\alpha_{10} b_1 |101>|101>+\alpha_{11} b_0 |110>|110>+\alpha_{11} b_1 |111>|111> $ \\
\hline
00 & 1 &  $\alpha_{00} b_0|001>|000>+\alpha_{00} b_1 |000>|001>+\alpha_{01} b_0 |011>|010>+\alpha_{01} b_1|010>|011> $ \\
& & $+\alpha_{10} b_0 |101>|100>+\alpha_{10} b_1 |100>|101>+\alpha_{11} b_0 |111>|110>+\alpha_{11} b_1 |110>|111> $ \\
\hline
 01 & 0 & $\alpha_{00} b_0|010>|000>+\alpha_{00} b_1 |011>|001>+\alpha_{01} b_0 |000>|010>+\alpha_{01} b_1|001>|011> $ \\
& & $+\alpha_{10} b_0 |110>|100>+\alpha_{10} b_1 |111>|101>+\alpha_{11} b_0 |100>|110>+\alpha_{11} b_1 |101>|111> $ \\
\hline
01 & 1 &  $\alpha_{00} b_0|011>|000>+\alpha_{00} b_1 |010>|001>+\alpha_{01} b_0 |001>|010>+\alpha_{01} b_1|000>|011> $ \\
& & $+\alpha_{10} b_0 |111>|100>+\alpha_{10} b_1 |110>|101>+\alpha_{11} b_0 |101>|110>+\alpha_{11} b_1 |100>|111> $ \\
\hline
10 & 0 &  $\alpha_{00} b_0|100>|000>+\alpha_{00} b_1 |101>|001>+\alpha_{01} b_0 |110>|010>+\alpha_{01} b_1|111>|011> $ \\
& & $+\alpha_{10} b_0 |000>|100>+\alpha_{10} b_1 |001>|101>+\alpha_{11} b_0 |010>|110>+\alpha_{11} b_1 |011>|111> $ \\
\hline
10 & 1 &  $\alpha_{00} b_0|101>|000>+\alpha_{00} b_1 |100>|001>+\alpha_{01} b_0 |111>|010>+\alpha_{01} b_1|110>|011> $ \\
& & $+\alpha_{10} b_0 |001>|100>+\alpha_{10} b_1 |000>|101>+\alpha_{11} b_0 |011>|110>+\alpha_{11} b_1 |010>|111> $ \\
\hline
11 & 0 &  $\alpha_{00} b_0|110>|000>+\alpha_{00} b_1 |111>|001>+\alpha_{01} b_0 |100>|010>+\alpha_{01} b_1|101>|011> $ \\
& & $+\alpha_{10} b_0 |010>|100>+\alpha_{10} b_1 |011>|101>+\alpha_{11} b_0 |000>|110>+\alpha_{11} b_1 |001>|111> $\\
\hline
11 & 1 &  $\alpha_{00} b_0|111>|000>+\alpha_{00} b_1 |110>|001>+\alpha_{01} b_0 |101>|010>+\alpha_{01} b_1|100>|011> $ \\
& & $+\alpha_{10} b_0 |011>|100>+\alpha_{10} b_1 |010>|101>+\alpha_{11} b_0 |001>|110>+\alpha_{11} b_1 |000>|111> $ \\
\hline
\end{tabular}
\end{table}
\textbf{Step 4:} The measurement results that obtained from previous step are notified by Alice and Bob to each other. Then, they apply the unitary  X operator  to their unmeasured qubits (Main q
qubits) according to the Table 2. As a result, the state of the whole system will be converted in the form as follow:
\begin{equation}
|\phi>_{(b0)(a0)(a1)(c_{a0})(c_{a1})(c_{b0})A0A1B0}=
\end{equation}
$$\alpha_{00} b_0|000>|000>+\alpha_{00} b_1 |001>
|001>+\alpha_{01} b_0 |010>|010>+\alpha_{01} b_1 |011>|011>$$ 
$$+\alpha_{10} b_0 |100>|100>+\alpha_{10} b_1 |101>|101>+\alpha_{11} b_0 |110>|110>+\alpha_{11} b_1 |111>|111>$$

\begin{table}[h!]
\caption{local unitary transformations performed by Alice and Bob}
\label{tab:1}
\begin{tabular}{|c|c|c|}
\hline
Alice’s Result	& Bob’s Result &	Unitary Operator on $(b0)(b1)(a0)$\\
\hline
00  & 0 & $I\otimes I\otimes I $\\
\hline
00  & 1 & $I\otimes I\otimes X $\\
\hline
01  & 0 & $I\otimes X\otimes I $\\
\hline
01  & 1 & $I\otimes X\otimes X $\\
\hline
10  & 0 & $X\otimes I\otimes I $\\
\hline
10  & 1 & $X\otimes I\otimes X $\\
\hline
11  & 0 & $X\otimes X\otimes I $\\
\hline
11  & 1 & $X\otimes X\otimes X $\\
\hline

\end{tabular}
\end{table}

\textbf{Step 5:} in this step, single-qubit measurements are applied on the Sending Qubits $(A_0,A_1$ and $B_0)$ by Alice and Bob  in the X-basis, respectively. As shown in the Table 3, the unmeasured qubits (Main qubits) collapse into the one of the 8 possible states with equal probability.

\begin{table}[h!]
\caption{measurement results based on X done by Alice and Bob on Main qubits}
\label{tab:1}
\begin{tabular}{|c|c|c|}
\hline
Alice's Result & Bob's Result & The collapsed state of qubits \\
 &  & $(b0)(b1)(a0)$\\
\hline
++ & + &  $\alpha_{00} b_0|000>+\alpha_{00} b_1 |001>+\alpha_{01} b_0 |010>+\alpha_{01} b_1 |011> $\\
& & $+\alpha_{10} b_0 |100>+\alpha_{10} b_1 |101>+\alpha_{11} b_0 |110>+\alpha_{11} b_1 
|111> $ \\
\hline
++ & - &   $\alpha_{00} b_0|000>-\alpha_{00} b_1 |001>+\alpha_{01} b_0 |010>-\alpha_{01} b_1 |011> $\\
& & $+\alpha_{10} b_0 |100>-\alpha_{10} b_1 |101>+\alpha_{11} b_0 |110>-\alpha_{11} b_1 
|111> $ \\
\hline
 +- & + &   $\alpha_{00} b_0|000>+\alpha_{00} b_1 |001>-\alpha_{01} b_0 |010>-\alpha_{01} b_1 |011> $\\
& & $+\alpha_{10} b_0 |100>+\alpha_{10} b_1 |101>-\alpha_{11} b_0 |110>-\alpha_{11} b_1 
|111> $ \\
\hline
+- & - &    $\alpha_{00} b_0|000>-\alpha_{00} b_1 |001>-\alpha_{01} b_0 |010>+\alpha_{01} b_1 |011> $\\
& & $+\alpha_{10} b_0 |100>-\alpha_{10} b_1 |101>-\alpha_{11} b_0 |110>+\alpha_{11} b_1 
|111> $ \\
\hline
-+ & + &   $\alpha_{00} b_0|000>+\alpha_{00} b_1 |001>+\alpha_{01} b_0 |010>+\alpha_{01} b_1 |011> $\\
& & $-\alpha_{10} b_0 |100>-\alpha_{10} b_1 |101>-\alpha_{11} b_0 |110>-\alpha_{11} b_1 
|111> $ \\
\hline
-+ & - &   $\alpha_{00} b_0|000>-\alpha_{00} b_1 |001>+\alpha_{01} b_0 |010>-\alpha_{01} b_1 |011> $\\
& & $-\alpha_{10} b_0 |100>+\alpha_{10} b_1 |101>-\alpha_{11} b_0 |110>+\alpha_{11} b_1 
|111> $ \\
\hline
- - & + &   $\alpha_{00} b_0|000>+\alpha_{00} b_1 |001>-\alpha_{01} b_0 |010>-\alpha_{01} b_1 |011> $\\
& & $-\alpha_{10} b_0 |100>-\alpha_{10} b_1 |101>+\alpha_{11} b_0 |110>+\alpha_{11} b_1 
|111> $ \\
\hline
- - &  - &   $\alpha_{00} b_0|000>-\alpha_{00} b_1 |001>-\alpha_{01} b_0 |010>+\alpha_{01} b_1 |011> $\\
& & $-\alpha_{10} b_0 |100>+\alpha_{10} b_1 |101>+\alpha_{11} b_0 |110>-\alpha_{11} b_1 
|111> $ \\
\hline
\end{tabular}
\end{table}

\textbf{Step 6:}After that users announced the results of the measurement to each other, they apply Z unitary operator to their unmeasured qubits (Main qubits) according to Table 4. Therefore, the state of the Main qubits will be converted as:
\begin{equation}
Main\;\; qubit=\alpha_{00} b_0|000>+\alpha_{00} b_1 |001> +\alpha_{01} b_0 |010> 
\end{equation}
 $$+\alpha_{01} b_1 |011>+\alpha_{10} b_0 |100>+\alpha_{10} b_1 |101>+\alpha_{11} b_0 |110>+\alpha_{11} b_1 |111>$$
 
 \begin{table}[h!]
\caption{local unitary transformations performed by Alice and Bob}
\label{tab:1}
\begin{tabular}{|c|c|c|}
\hline
Alice’s Result	& Bob’s Result &	Unitary Operation on $(b0)(b1)(a0)$\\
\hline
++  & + & $I\otimes I\otimes I $\\
\hline
++  & - & $I\otimes I\otimes Z $\\
\hline
+ -  & + & $I\otimes Z\otimes I $\\
\hline
+ -  & - & $I\otimes Z\otimes Z $\\
\hline
- +  & + & $ Z\otimes I\otimes I $\\
\hline
- +  & - & $Z\otimes I\otimes Z $\\
\hline
- -  & + & $Z\otimes Z\otimes I $\\
\hline
- -  & - & $Z\otimes Z\otimes Z $\\
\hline

\end{tabular}
\end{table}

\textbf{Step 7:} Alice and Bob reconstruct the one and two-qubit states according to the bellow equation and the BQT is successfully finished.
$$
|\phi>_{A0}=(b_0|0>+b_1|1>)_{a0}
$$
\begin{equation}
|\phi>_{B0B1}=(\alpha_{00}|00>+\alpha_{01}|01>+\alpha_{10}|10>+\alpha_{11}|11>)_{b0b1} 
\end{equation}

\subsection{Transmitting of n and m qubits by Alice and Bob with entanglement state}
 Suppose that one of the states or each two states transmitted by Alice and Bob are entangled. In this case that the state is entanglement, only the Step 1 of the protocol (channel structure) is changed. For example, suppose Alice wants transmit n-qubit entanglement state to Bob and Bob wants transmit a m-qubit state to Alice as shown in below equation (and Fig. 2 (a))
 \begin{equation}
 |\phi>_{an}=\alpha_{0}|0...0>+\alpha_{1}|1...1> \;\; \alpha_{0},\alpha_{1}\in C_{z}\;\; ,|\alpha_{0}|^2+|\alpha_{1}|^2=1 
\end{equation}  
\begin{equation}
|\phi>_{bm}=\sum_{i=0}^{2^{m}-1}b_{i}|i> \;\; b_{i}\in C_{z}\;\; ,\sum_{i=0}^{2^{n}-1}b_{i}^{2}=1
\end{equation}

\begin{figure}[h!]
\center
 (a)\includegraphics[width=0.5\textwidth]{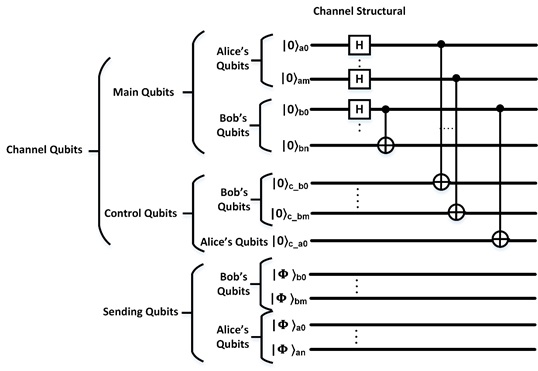}
  (b)\includegraphics[width=0.5\textwidth]{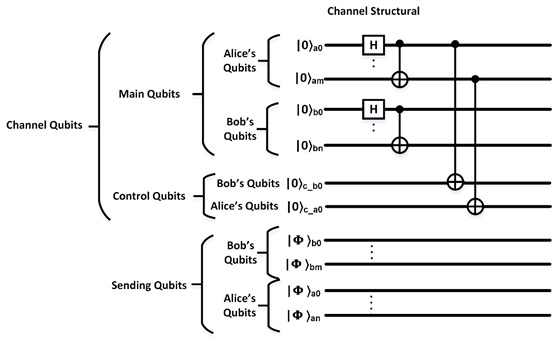}
 \caption{Channel structural. a)The n-qubit entanglement state as a sending qubits by Alice. b)The n and m qubits entanglement state as a sending qubits by Alice and Bob.}
 \label{1}
\end{figure}
 
 As shown in Fig. 2 (a), Main qubit is constructed from m-qubit for Alice and n-qubit for Bob as $|0>_{a0}...|0>_{am}|0>_{b0}...|0>_{bn}$. This n-qubit has been supposed as entanglement state. So, in the Main qubit, the operator that introduced in previous section applied on the Alice's m-qubit and  for n-qubit of Bob, the following operator is applied.  Hadamard gate applied on the first qubit then CNOT operator applied. So that the first qubit is considered as control lines and other qubits as target lines.\\
 For control qubits of $|0>_{(c_{b0})}...|0>_{(c_{bm})}$ , CNOT gates are applied as described in the previous section. But, the number of control-qubits for a n-qubit entanglement state is one. So, a CNOT gate is applied to the state of  $|0>_{(c_{a0})}$. So that, the first qubit of Bob in main qubits is considered as control line and $c_{a0}$ line is as target line.\\
  
Fig. 2 (b) shows conditions that Alice and Bob transmit $n$ and $m$ qubit entanglement states to each other. So, it need to two control qubits and structure implemented for Alice in Fig. 2 (a) is considered for each of Alice and Bob. The other steps of protocol are same as before. Only in the Step 2 of protocol, for applying CNOT gates to Control qubits, each of Sending qubits by Alice and Bob are as entanglement state. So, the first it's qubit only is considered as control line for related Control qubits.
\\

\textbf{Example 2:} consider a two-qubit entanglement state that Alice wants teleport to Bob and a one-qubit state that Bob wants to teleport to Alice are given by :
\begin{equation}
|\phi>_{A0A1}=\alpha_{0}|00>+\alpha_{1}|11> 
\end{equation}
\begin{equation}
|\phi>_{B}=b_0|0>+b_1|1> 	
\end{equation}

For creating the channel, two qubits are considered for Bob and one qubit is considered for Alice. Also, one qubit for Alice and one qubit for Bob as Control Qubits are considered. Then, the general state is as the following:
\begin{equation}
|000>_{b0b1a0} |00>_{(c_a)(c_b)}
\end{equation}

Then, for main qubits Hadamard gate is applied to the first Bob’s qubit (because it is entanglement) and Alice qubits $(a0)$. The created state is 
\begin{equation}
(|000>+|001>+|100>+|101>)_{b0b1a0} |00>_{(c_a)(c_b)} 
\end{equation}
Then, CNOT gates are applied to Bob's qubits in the Main qubits so that, the first Bob's qubit $(b0)$ is considered as control line and other qubits $(b1)$ belong to Bob are considered as target lines. Therefore the created state is as following:
\begin{equation}
\frac{1}{2}(|000>+|001>+|110>+|111>)_{b0b1a0}|00>_{(c_a)(c_b)} 
\end{equation}
Then, for control qubits CNOT gates are applied so that, b0 and a0 are considered as control lines. $c_a$ and $c_b$ are considered as target lines. Therefore, the created channel is as following: 
\begin{equation}
|\psi>_{(b0)(b1)(a0)(c_a)(c_b)} =\frac{1}{2}(|000>|00>+|001>|01>+|110>|10>+|111>|11>)
\end{equation}
The general state of the system is 
\begin{equation}
\mid\phi>_{(b0)(b1)(a0)(c_a)(c_b)A0A1B0}=|\psi>_{(b0)(b1)(a0)(c_a)(c_b)}\otimes |\phi>_{(A_0 A_1 )}\otimes|\phi_{(B_0)}
\end{equation}
For Step 2 of this protocol, CNOT operators are applied by Alice and Bob so that, $A_0$ and $B_0$ are as control lines and $c_a$ and $c_b$ are as target lines, respectively. The other Steps of the protocol are same as with previous protocol.

\subsection{Bidirectional Quantum Controlled Teleportation}
For transmitting $n $and $m$ qubit by Alice and Bob to each other هد the presence of a controller Charlie, the proposed methods in the previous section can be used but, a qubit belong to Charlie is added to the last Main qubits (Placing of Charlie's qubit is arbitrary). Also, after that all previous operators applying in creating channels, only  CNOT gates is applied on it. Therefore each combination of Control Qubits can be considered as control lines and Charlie's qubit is as target line. So, Charlie can create $2^{(the number of Control Qubits)}$ quantum channel that in each quantum channel, the distribution of  charlie's qubit is different. Then Charlie can encode each type of channel with classical bit and notify the type of creating a channel to Alice and Bob in the final step. Thus by doing this, the probability of estimation of Charlie's qubit by eavesdropper is $\frac{1}{2^{(the number of Control Qubits)}}$ . For example, channel created in Example 1 can be converted to a bidirectional controlled quantum teleportation protocol, thus the first, initial state is created as the following:
\begin{equation}
|0000>_{b0b1a0C}|000>_{(c_{a0} c_{a1} c_{b0} )}
\end{equation}
Where, $C$ denote belong to Charlie. Then, Applying of Hadamard gate to main qubits have:
\begin{equation}
\frac{1}{\sqrt{8}}(|0000>+|0010>+|0100>+|0110>+|1000>+|1010>
\end{equation}
$$
+|1100>+|1110>)_{b0b1a0C}(|000>)_{(c_{a0} c_{a1} c_{b0})}
$$	

Then, by applying CNOT gate to control qubits we have:
\begin{equation}
 \frac{1}{\sqrt{8}}((|0000>|000>+|0010>|001>+|0100>|010>+|0110>|011> 
 \end{equation}
$$ 
+|1000>|100>+|1010>|101>+|1100>|110>+|1110>|111>)_{((b0)(b1)(a0)(C)(c_{a0})(c_{a1})(c_{b0}) )} 
$$
Finally, a CNOT gate is applied to Charlie's qubit (C) according to combination of Control qubits as control lines and Charlie's qubit as target line. So, channels are created as shown in Table 5.

\begin{table}[h!]
\caption{measurement results based on X done by Alice and Bob on Main Qubits}
\label{tab:1}
\begin{tabular}{|c|c|c|c|}
\hline
Encoding of the & The proposed channel  &Charlie’s & Unitary \\
various distributions  & $|\Psi>_{(b_0)(b_1)(a_0)(C)(Ca_0)(C_a1)(C_b0)}$ & Results & Operation \\
of Charlie’s qubit &	& 	&on $b_0,b_1,a_0$\\
\hline
000 & $\frac{1}{\sqrt{8}}(|0000>|000>+|0010>|001> $ & $|+>$ & $I\otimes I\otimes I$\\
 & $+|0100>|010>+|0110>|011>+|1000>|100>$ & $|->$ &$I\otimes I\otimes I$\\
 &$+|1010>|101>+|1100>|110>+|1110>|111>)$ & &\\
 \hline
 001 & $\frac{1}{\sqrt{8}}(|0000>|000>+|0011>|001> $ & $|+>$ & $I\otimes I\otimes I$\\
 & $+|0100>|010>+|0111>|011>+|1000>|100>$ & $|->$ &$I\otimes I\otimes Z$\\
 &$+|1011>|101>+|1100>|110>+|1111>|111>)$ & &\\
 \hline
010 & $\frac{1}{\sqrt{8}}(|0000>|000>+|0010>|001> $ & $|+>$ & $I\otimes I\otimes I$\\
 & $+|0101>|010>+|0111>|011>+|1000>|100>$ & $|->$ &$I\otimes Z\otimes I$\\
 &$+|1010>|101>+|1101>|110>+|1111>|111>)$ & &\\
 \hline
 011 & $\frac{1}{\sqrt{8}}(|0000>|000>+|0011>|001> $ & $|+>$ & $I\otimes I\otimes I$\\
 & $+|0101>|010>+|0110>|011>+|1000>|100>$ & $|->$ &$I\otimes Z\otimes Z$\\
 &$+|1011>|101>+|1101>|110>+|1110>|111>)$ & &\\
 \hline
 100 & $\frac{1}{\sqrt{8}}(|0000>|000>+|0010>|001> $ & $|+>$ & $I\otimes I\otimes I$\\
 & $+|0100>|010>+|0110>|011>+|1001>|100>$ & $|->$ &$Z\otimes I\otimes I$\\
 &$+|1011>|101>+|1101>|110>+|1111>|111>)$ & &\\
 \hline
 101 & $\frac{1}{\sqrt{8}}(|0000>|000>+|0011>|001> $ & $|+>$ & $I\otimes I\otimes I$\\
 & $+|0100>|010>+|0111>|011>+|1001>|100>$ & $|->$ &$Z\otimes I\otimes Z$\\
 &$+|1010>|101>+|1101>|110>+|1110>|111>)$ & &\\
 \hline
 110 & $\frac{1}{\sqrt{8}}(|0000>|000>+|0010>|001> $ & $|+>$ & $I\otimes I\otimes I$\\
 & $+|0101>|010>+|0111>|011>+|1001>|100>$ & $|->$ &$Z\otimes Z\otimes I$\\
 &$+|1011>|101>+|1100>|110>+|1110>|111>)$ & &\\
 \hline
 111 & $\frac{1}{\sqrt{8}}(|0000>|000>+|0011>|001> $ & $|+>$ & $I\otimes I\otimes I$\\
 & $+|0101>|010>+|0110>|011>+|1001>|100>$ & $|->$ &$Z\otimes Z\otimes Z$\\
 &$+|1010>|101>+|1100>|110>+|1111>|111>)$ & &\\
 \hline
\end{tabular}
\end{table}

 Suppose channel created by the controller (Charlie) be encoded channel by classical bits 001 shown as 
 \begin{equation}
|\Psi>_{((b0)(b1)(a0)(C)(c_{a0})(c_{a1})(c_{b0}))}=\frac{1}{\sqrt{8}}((|0000>|000>+|0011>|001>+|0100>|010>
\end{equation}

$$
+|0111>|011>+|0100>|010>+|0111>|011>+|0111>|011>
$$
$$
+|1000>|100>+|1011>|101>+|1100>|110>+|1111>|111>)
$$
Then, the other steps of the protocol are same as the proposed protocol in previous sections. Only, after than the Step 6, Charlie must allow Bob and Alice for reconstructing the initial unknown state (Equation that written in below shows the wholes state of the system after than Step 6.). Thus, he needs to apply single qubit measurement in the X-basis on its qubit and tells to receivers his result and also, he must tell them about creating a channel by encoding bits. Therefore, we create Step 6-1 and Step 6-2 for it as the following:
\begin{equation}
(\alpha_{00} b_0 |0000>+\alpha_{00} b_1 |0011>+\alpha_{01} b_0 |0100>+\alpha_{01} b_1|0111>+\alpha_{10} b_0 |1000>
\end{equation}
$$
+\alpha_{10} b_1 |1011>+\alpha_{11} b_0 |1100>+\alpha_{11} b_1 |1111>)_{(b0)(b1)(a0)(C)}
$$
\textbf{Step 6-1:} For reconstructing the initial state by Alice and Bob, Charlie must allow them by measuring in the X-basis on its qubit and tells to Alice and Bob his result. Also, he tells the type of created channel to them by encoding bits 001. If Charlie’s measured result is $|+> (|->)$, then the state of other qubits is 
\begin{equation}
(\alpha_{00} b_0 |000>+\alpha_{00} b_1 |001>+\alpha_{01} b_0 |010>+\alpha_{01} b_1 |011>+
\end{equation}
$$\alpha_{10} b_0 |100>+\alpha_{10} b_1 |101>+\alpha_{11} b_0 |110>+\alpha_{11} b_1 |111>)$$ 	
\\
\begin{equation}
(\alpha_{00} b_0 |000>-\alpha_{00} b_1 |001>+\alpha_{01} b_0 |010>-\alpha_{01} b_1 |011>+
\end{equation}
$$
\alpha_{10} b_0 |100>-\alpha_{10} b_1 |101>+\alpha_{11} b_0 |110>-\alpha_{11} b_1|111>)
$$ 

\textbf{Step 6-2:} After than Charlie tells its measurement result and type of the created channel to Alice and Bob. Then, they need to apply local unitary operator as showed in the last column of Table 5.\\
Then, In Step 7, Alice and Bob can reconstruct transmitted states again and the BQCT is successfully finished.

\section{Comparison} 
In this section, a comparison between quantum channels introduced in previous works and our proposed method for creating BQT protocol are presented as shown in Table 6 for many of the best previous works with various Sending qubits. In this table, columns of NUM Alice and NUM Bob stand for numbers of transmitted qubits between Alice and Bob, respectively. Also, Channel column shows introduced channel by previous works and our method. In addition, final column express analogy between channel created by our method and previous works. As shown in this table, many of channels introduced in the previous works are same as with channels in proposed method with difference in moving of Control qubits or Charlie’s qubit \cite{Hong,Hassanpour2,Li3,Deng2} where as stated in Section 1, placing of these qubits are arbitrary and can place everywhere. But, in many of previous works \cite{Chen,Sang}, channel created by previous works are different from our method. These differences are due to that in these works two-qubit measurement are used such as measurement based on Bell state while in our method one-qubit measurement is used for creating channel that it is more efficient from previous works. Also, our method  is a protocol with the numbers of Fewer qubits in channel than many previous works such as \cite{Chen,Hassanpour}.

\section{Conclusion}
In this paper, we presented a new general protocol of Bidirectional Quantum Teleportation (BQT) and Bidirectional Quantum Controlled Teleportation (BQCT) to  transmit the arbitrary numbers of qubits between Alice and Bob with or without supervisor Charlie. This method can cover transmitting of each arbitrary types of qubits such as entanglement states or pure states simultaneously. The users can teleport each type state by using a $2×(n+m)$-qubit channel for  the not entanglement states(pure states), $(n+2m+1)$-qubit channel  when one of the states  is entanglement (for exp. State belong to Alice) and $(n+m+2)$-qubit channel for  transmitting the entanglement states. Where n and m denote the number of Sending qubits between Alice and Bob, respectively. Also, for BQCT protocol, due to presence of Charlie, one qubit is added to the number of Main qubits. For creating channels, CNOT operator and Hadamard gate are used. In addition, this protocol uses single-qubit measurement based on Z or X which are more efficient than two-qubit measurements \cite{Deng2}. We hope that such BQT and BQCT protocols can be realized experimentally in the future. 

\newpage

\pagestyle{empty}

\begin{landscape}
\begin{table}[h!]
\scriptsize
\caption{measurement results based on X done by Alice and Bob on Main qubits}
\label{tab:1}
\begin{tabular}{|c|c|c|c|c|c|}
\hline
Reference&Protocol&NUM&NUM&Channel&Analogy between proposed channel\\
&type&Alice&Bob& &and previous one \\
\hline
[33] method & BQT&2&2&$\frac{1}{4}[|00000000>+|00010001>+|00100010>+|00110011>+|01000100>$&Moving of Control Qubits\\
&&&&$+|01010101>+|01100110>+|01110111>+|10001000>+|10011001>$&\\
&&&&$+|10101010>+|10111011>+|11001100>+|11011101>+|11101110>$&\\
&&&&$+|11111111>]_{a1b1a2b2b3b4a3a4}$&\\
\hline
our method&BQT&2&2&$\frac{1}{4}[|00000000>+|00010001>+|00100010>+|00110011>+|01000100>$&Moving of Control Qubits\\
&&&&$+|01010101>+|01100110>+|01110111>+|10001000>+|10011001>$&\\
&&&&$+|10101010>+|10111011>+|11001100>+|11011101>+|11101110>$&\\
&&&&$+|11111111>]_{b0b1a0a1Ca0Ca1Cb0Cb1}$&\\
\hline
[32] method &BCQT&2 entanglement&1&$\frac{1}{2}[|000000>+|000111>+|111000>-|111111>]_{123456}$&Moving of control qubits\\
&&&&& and Charlie qubits\\
\hline
our method &BCQT&2 entanglement&1&$\frac{1}{2}[|000000>+|001101>+|110010>+|111111>]_{b0b1a0cca0cb0}$&Moving of Control Qubits\\
&&&&& and Charlie qubits\\
\hline
[31] method&BQT&2 entanglement&2 entanglement&$\frac{1}{2}[|000000>+|000111>+|111000>+|111111>]_{a1b1b2a2a3b3}$&Moving of Control Qubits\\
\hline
our method&BQT&2 entanglement&2 entanglement&$\frac{1}{2}[|000000>+|001101>+|110010>+|111111>]_{b0b1a0a1ca0cb0}$&Moving of Control Qubits\\
\hline
[30] method&BCQT& 1 &2 &$\frac{\sqrt{2}}{4}[|0000000>+|0000011>+|0001101>+|0001110>+|1110001>$&Moving of Control Qubits\\
&&&&$+|1110010>+|1111100>+|1111111>]_{1234567}$\\
\hline
our method&BCQT&1 &2 &$\frac{\sqrt{2}}{4}[|0000000>+|0011001>+|0100010>+|0111011>+|1000100>$&Moving of Control Qubits\\
&&&&$+|1011101>+|1100110>+|1111111>]_{b0a0a1cca0cb0cb1}$\\
\hline
[29] method&BCQT& 1 &2 &$\frac{\sqrt{2}}{4}[|0000000>+|0010010>+|0100100>+|0110110>+|1001001>$&Moving of Charli Qubits and difference in variations of\\
&&&&$+|1011011>+|1101101>+|1111111>]_{1234567}$& Charlie's Qubits this qubit change according to qubot b0 \\
\hline
our method&BCQT&1&2 &$\frac{\sqrt{2}}{4}[|0000000>+|0011001>+|0100010>+|0111011>+|1000100>$&in our method change according qubit cb1\\
&&&&$+|1011101>+|1100110>+|1111111>]_{b0a0a1cca0cb0cb1}$\\
\hline
[28] method&BCQT& 1 &1 &$\frac{1}{2 \sqrt{2}}[|000000>+|000011>+|001100>+|001111>+|110000>$&Moving of Contorol Qubits by 6 qubit channel\\
&&&&$+|110011>+|111100>+|111111>]_{a1b1c1a2c2b2}$&  \\
\hline
our method&BCQT&1&1 &$\frac{1}{2}[|00000>+|01101>+|10010>+|11111>]_{b0a0cca0cb0}$&in our method 5 qubit channel used\\
\hline
[27] method&BCQT& 1 &1 &$\frac{\sqrt{2}}{4}[|000000>+|011100>+|111000>+|100100>+|001111>$&Moving of Contorol Qubits by 6 qubit channel\\
&&&&$+|010011>+|110111>+|101011>]_{123456}$&  \\
\hline
our method&BCQT&1&1 &$\frac{1}{2}[|00000>+|01101>+|10010>+|11111>]_{b0a0cca0cb0}$&in our method 5 qubit channel used\\
\hline

\end{tabular}
\end{table}
\end{landscape}

\pagestyle{plain}

\section{Acknowledgement}
This work is supported by Kermanshah Branch, Islamic Azad
University, Kermanshah, IRAN.

\end{document}